\documentclass[epj-spec]{svjour}
\usepackage{graphics}
\usepackage{epsf,epsfig}

\def\xb{\overline{x}}

\def\als{\alpha_s}
\def\vk{{\bf k}_{\perp}}
\def\vbs{{\bf b}}

\begin{document}
\title{Effects of quark GPDs in   meson electroproduction.}
\author{S.V.Goloskokov \thanks{\email{goloskkv@theor.jinr.ru}} }
\institute{ Bogoliubov Laboratory of Theoretical  Physics,
  Joint Institute for Nuclear Research, Dubna, Russia}
\abstract{We analyze light  meson electroproduction  in the
intermediate energy range where the quark contributions are
essential. Our calculations are carried out on the basis of the
handbag approach. We study cross sections and spin asymmetries for
various vector and pseudoscalar mesons. Our results  are in good
agrement with   COMPASS and HERMES experiments.
} 
\maketitle

 \section{Introduction}
This report is devoted to the investigation of spin effects in
hard electroproduction of vector and pseudoscalar mesons off
unpolarized and transversally polarized protons. In the kinematic
region of interest the leading twist amplitude factorizes
\cite{fact} into hard meson electroproduction off partons and
generalized parton distribution GPD. GPDs depend on  $\bar x$ -the
momentum fraction of the parton, $\xi$- skewness which is
determined as a difference of the parton momenta and $t$-momentum
transfer. The skewness $\xi$ is related to Bjorken- $x_B$ as $\xi
\sim x_B/2$

GPDs contain the extensive information on the hadron structure. At
zero skewness and momentum transfer GPD become equal to the
corresponding parton distribution PDF. The form factors of hadron
can be calculated from GPDs trough the integration over $\bar x$.
Using Ji sum rules \cite{ji} the parton angular momentum can be
extracted.

Our calculations \cite{gk05,gk06,gk07q} are based on the approach
where we modify the collinear approximation by including the quark
transverse degrees of freedom and the Sudakov corrections in the
hard subprocess amplitude. This regularizes singularities in the
higher twist TT amplitude which is essential in the description of
spin effects. Within our approach we calculate the process
amplitudes and, subsequently, the cross sections and the spin
observables in the light meson electroproduction. Using GPDs $E$
we extend our analysis to a transversally polarized target. We
calculate cross sections and the $A_{UT}$ asymmetry for various
vector meson productions \cite{gk08}. The study of pion
electroproduction \cite{gk09} presented here gives access to the
GPD $\tilde H$ and $\tilde E$.

Our results \cite{gk06,gk07q,gk08,gk09} on meson electroproducion
at moderate $x$ are in good agreement with experimental data at
COMPASS \cite{compass} and HERMES \cite{hermes} energy range.

\section{The handbag approach to  meson Electroproduction}
Here we are interested mainly in moderate HERMES and COMPASS
energies where the quark contributions are essential. From
different meson productions  we can get information about valence
and sea quark effects. Really, the quarks contribute to meson
production reactions in different combinations. For uncharged
meson production we have standard GPDs and find
\begin{equation}\label{quarks}
\rho:\;\; \propto \frac{2}{3} H^u +\frac{1}{3} H^d;\;\;\omega:\;\;
\propto \frac{2}{3} H^u -\frac{1}{3} H^d.
\end{equation}
For production of charged and strange mesons we have transition $
p\to n$ and $p\to \Sigma$ GPDs:
\begin{equation}\label{quarkspl}
\rho^+:\;\; \propto H^u -H^d;\;\;\;\;K^{*0}:\;\;  \propto  H^d
-H^s.
\end{equation}
For pseudoscalar mesons the polarized GPDs contribute to
\begin{equation}
 \pi^+:\;\;\propto \tilde H^u -
\tilde H^d ;\;\;\pi^0:\;\; \propto \frac{2}{3} \tilde H^u
+\frac{1}{3} \tilde H^d.
\end{equation}
The same combinations are valid for $E, \tilde E$ contributions.
Thus, we can test various GPDs  in the mentioned reactions.

Unfortunately, the amplitudes of meson production contain GPD
 in complicated integrated forms and direct extraction
of GPDs from amplitudes is impossible. So we parameterize GPDs
from model approach calculate process amplitudes and compare
results with experiment. To estimate  GPDs, we use the double
distribution representation \cite{mus99}
\begin{equation}
  H_i(\xb,\xi,t) =  \int_{-1}
     ^{1}\, d\beta \int_{-1+|\beta|}
     ^{1-|\beta|}\, d\alpha \delta(\beta+ \xi \, \alpha - \xb)
\, f_i(\beta,\alpha,t)
\end{equation}
which connects  GPDs with PDFs through the double distribution
function $f$,

\begin{equation}\label{ddf}
f_i(\beta,\alpha,t)= e^{b_i\,t}\, |\beta|^{-\alpha't}\,
h_i(\beta)\,
                   \frac{\Gamma(2n_i+2)}{2^{2n_i+1}\,\Gamma^2(n_i+1)}
                   \,\frac{[(1-|\beta|)^2-\alpha^2]^{n_i}}
                           {(1-|\beta|)^{2n_i+1}}.
                          \end{equation}
The functions $h_i(\beta)$  are expressed in terms of PDFs. In
(\ref{ddf}) the Regge motivated ansatz is used to model $t$
dependences of PDFs. We assume that the Regge trajectories are
linear functions of $t$ at small momentum transfer
$\alpha_i=\alpha_i(0)+\alpha'_i t$. The powers in (\ref{ddf}) are
equal to n=2 for  gluon and sea and to n=1 for valence
contributions.

 To calculate GPDs, we use the CTEQ6 fits of  PDFs for gluon, valence
 and sea quarks \cite{CTEQ}. Note that the $u$ and $d$  seas have
very similar $\beta$ and $Q^2$ dependences. This is not the case
for the strange sea. In agrement with CTEQ6 PDFs  we suppose the
flavor asymmetric quark sea to be $H^u_{sea} = H^d_{sea} =
\kappa_s H^s_{sea}$ \cite{gk07q} with $\kappa_s$ determined from
CTEQ6.

The  hard subprocesses are analysed within the modified
perturbative approach MPA \cite{sterman} where we consider the
quark transverse degrees of freedom accompanied by Sudakov
suppressions. In the model, the amplitude of the  meson production
off the proton  reads as a convolution of the hard partonic
subprocess
 ${\cal H}$ and $\hat H_i$
\begin{equation}\label{amptt}
  {\cal M}^{a}_{\mu'\pm,\mu +} = \, \sum_{a}\, \langle {H}^a
  \rangle+  \langle \tilde{H}^a
  \rangle\;\;\;\;
\langle {H}^a\rangle = \sum_{\lambda}
         \int_{xi}^1 d\xb
        {\cal H}^{a}_{\mu'\lambda,\mu \lambda}(Q^2,\xb,\xi,t)
                                   \hat H^{a}(\xb,\xi,t)
\end{equation}
where  $a$ denotes the gluon and quark contribution with the
corresponding flavors \cite{gk06};
 $\mu$ ($\mu'$) is the helicity of the photon (meson), and $\xb$
 is the momentum fraction of the parton with helicity $\lambda$.

 The $\hat H^{a}$ in (\ref{amptt}) is expressed in terms of GPDs.
For the proton helicity nonflip  $\hat
H^{a}=[H^{a}-\frac{\xi^2}{1-\xi^2} E^{a}]$ which is close to
$H^{a}$. For the amplitude $M_{\mu' -,\mu +}$ with proton
helicity-flip in (\ref{amptt})  one should use $\hat
H^{a}=\sqrt{-t'}/(2 m) [E^{a} + \xi \tilde E^{a}]$.

 The  amplitude ${\cal H}^{a}$ is represented as  the contraction of the hard
  part, which is calculated perturbatively, and the
non-perturbative meson  wave function $ \phi_V$
\begin{equation}\label{hsaml}
  {\cal H}^{a}_{\mu'+,\mu +}\,=
\,\frac{2\pi \als(\mu_R)}
           {\sqrt{2N_c}} \,\int_0^1 d\tau\,\int \frac{d^{\,2} \vk}{16\pi^3}
            \phi_{V}(\tau,k^2_\perp)\;
                f^{a}_{\vk,\mu',\mu}(\xb,\xi,\tau) \hat{D} (\tau,Q,\vk).
\end{equation}
Here $\phi_{V}$ is a meson wave function, $f^a_{\mu',\mu}$ is a
hard subprocess amplitude and $\hat{D}$ is a denominator of
propagators in the hard amplitude. We keep the $k^2_\perp$ terms
in the denominators of LL and TT transitions and in the numerator
of the TT amplitude. The hard propagators now look like
\begin{equation}\label{propk}
\hat{D_i}=\frac{1}{d_i Q^2+k_\perp^2}
\end{equation}
and a singularity at the $d_i=0$ point disappears in (\ref{propk})
if $k_\perp^2 \neq 0$. This regularizes the higher twist $TT$
amplitude which is divergent at the collinear $\vk=0$
approximation. The hard amplitude is calculated using the $k$-
dependent wave function \cite{koerner} that contains the leading
and higher twist terms describing the longitudinally and
transversally polarized vector mesons, respectively \cite{gk07q}.
The wave function is chosen in the Gaussian form
\begin{equation}\label{wave-l}
  \phi_V(\vk,\tau)\,\propto f_V a^2_V
       \, \exp{\left[-a^2_V\, \frac{\vk^{\,2}}{\tau\bar{\tau}}\right]}\,.
\end{equation}
 Here $\bar{\tau}=1-\tau$, $f_V$ is the
decay coupling constant. The $a_V$ parameter determines the mean
value of the quark transverse momentum $<\vk^{\,2}>$  in the
vector meson. The $a_V$ values are  different for longitudinal and
transverse polarization of the meson \cite{gk05,gk06}.

  The gluonic
corrections are treated in the form of the Sudakov factors. The
resummation and exponentiation of the Sudakov corrections can be
performed  in the impact parameter space. We use the Fourier
transformation to transfer integrals from $\vk$ to $\vbs$ space.

\section{Vector meson electroproduction}
In this section, we  analyze vector meson electroproduction and
consider the cross section and spin observables. In the previous
section, we present the main ingredients that are needed for the
GPD and amplitude calculation. All GPDs are modeled on the basis
of the double distribution ansatz  with using CTEQ6
parameterization of PDFs. The $a_V$ parameters in the wave
function were determined from the best description of the cross
section. Some more details together with other parameters of the
model can be found in \cite{gk05,gk06,gk07q}.

 This approach has been found to be successful in the analysis of
data on $\rho^0$ and $\phi$ electroproduction. We consider the
gluon, sea and quark GPD contribution to the amplitude. This
permits us to analyze vector meson production \cite{gk06,gk07q} at
moderate values of $x$ ($ \sim 0.2$) typical of HERMES and
COMPASS. It was found that the $k_\perp^2/Q^2$ corrections
(\ref{propk}) in the hard amplitude are extremely important at low
$Q^2$. They decrease the cross section by a factor of about 10 at
$Q^2 \sim 3\mbox{GeV}^2$. Moreover, $k_\perp^2$ terms give a
possibility to calculate within our model the higher twist $TT$
amplitude which is responsible for the spin effects. The obtained
results  are in reasonable agreement with experiments at HERA
\cite{h1,zeus}, COMPASS \cite{compass}, HERMES \cite{hermes} and
E665 \cite{e665} energies for electroproduced $\rho$ and $\phi$
mesons \cite{gk06,gk07q}.

We have analyzed two different sets of the  spin density matrix
elements (SDME). The first set is  sensitive to the absolute
values of LL and TT transition amplitudes, the second one  is
expressed in terms of the LL and TT amplitude interference. The
description of experimental data for the first set of SDME is
quite good. For the second set of SDME that are relevant to the
phase shift  $\delta_{LT}$ between the LL and TT amplitudes the
situation is different. The HERMES experiment show a quite large
phase difference $\delta_{LT} \sim 20-30^o$; \cite{hermes};
however,  our model  gives small $\delta_{LT} \sim 2-3^o$
\cite{gk05}. A similar phase shift $\delta_{LT} \sim 20^o$ is
observed in H1 experiment \cite{h109}. This shows that such a
phase shift should appear in the  gluon contribution to the LL and
TT amplitudes and is not understood till now.

\begin{figure}[h!]
\begin{center}
\begin{tabular}{cc}
\includegraphics[width=6.8cm,height=4.9cm]{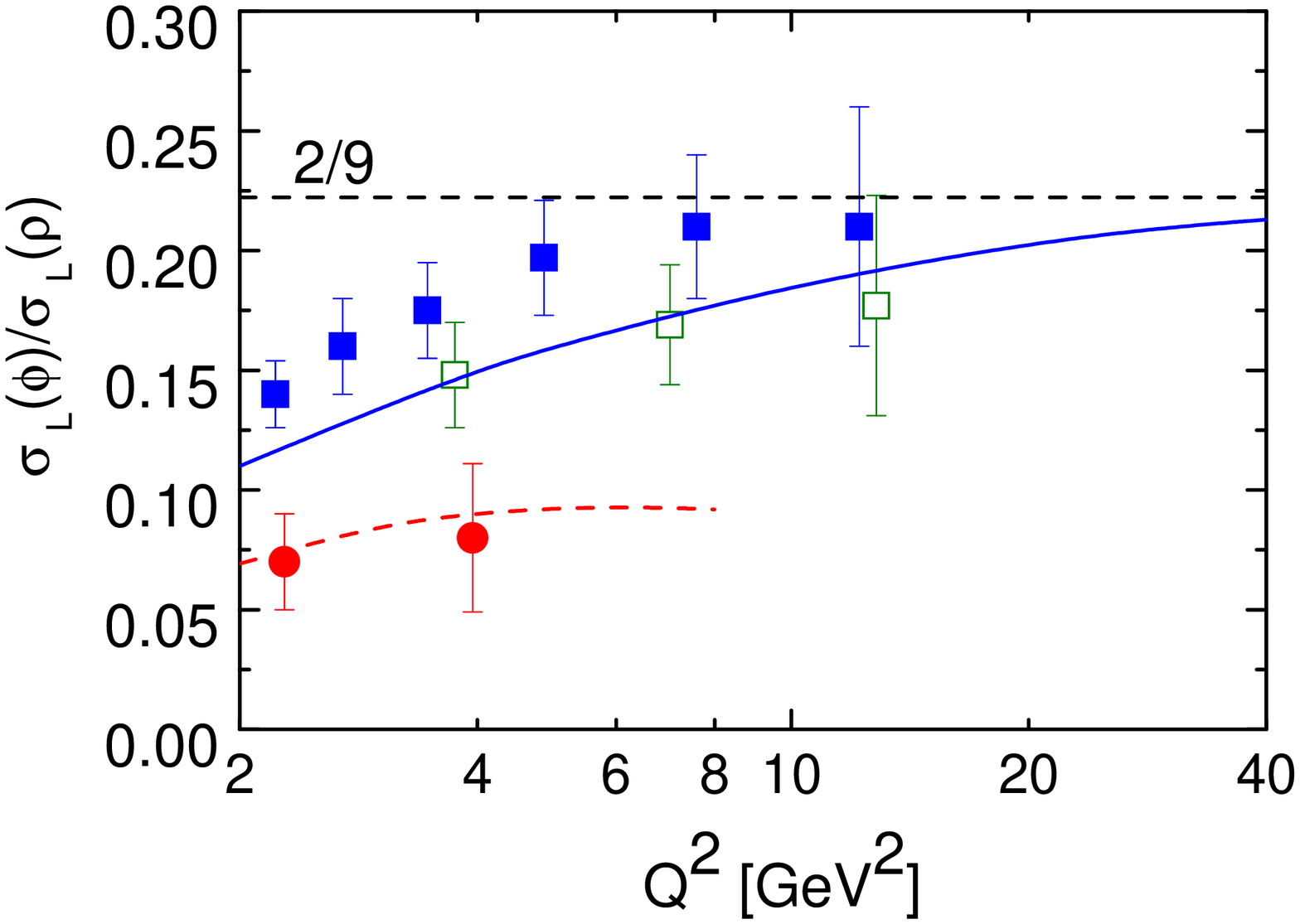}&
\includegraphics[width=6.8cm,height=4.9cm]{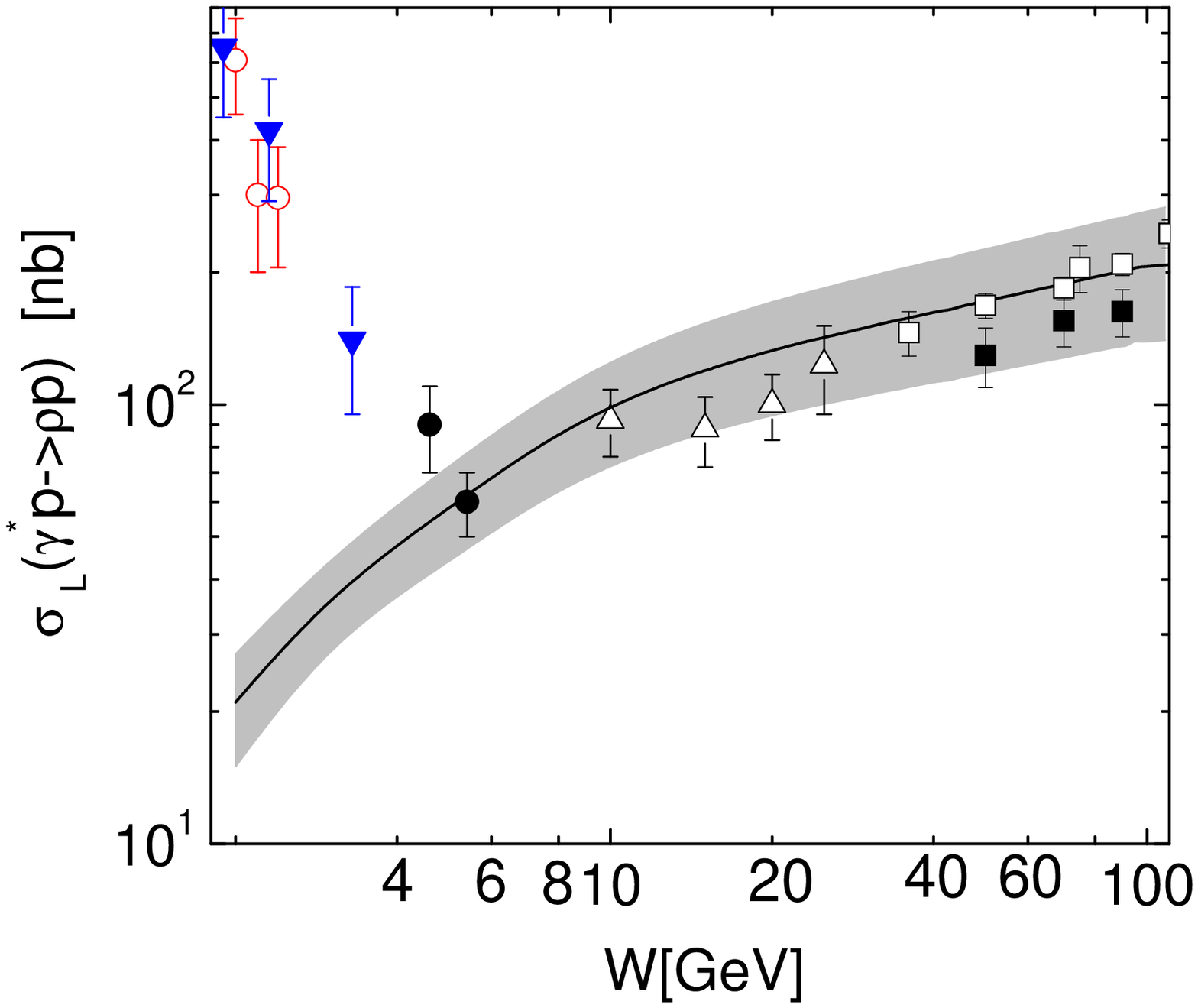}\\
{\bf(a)}& {\bf(b)}
\end{tabular}
\end{center}
\caption{ {\bf(a)} The  ratio of longitudinal cross sections
$\sigma_\phi/\sigma_\rho$ at
    HERA energies- full line and HERMES- dashed line. Data are from H1  -solid, ZEUS
 -open squares, HERMES solid circles. {\bf(b)}The longitudinal cross section for
 $\rho$ production at
$Q^2=4\,\mbox{GeV}^2$. Data: HERMES, ZEUS, H1, open circle- CLAS
data point.}
\end{figure}

The $Q^2$ dependence of the ratio $\sigma_\phi/\sigma_\rho$ cross
section at HERA and HERMES energies was investigated. At high
energies, where the valence quark effects are unessential, this
ratio for  the flavor symmetric sea   should be close to 2/9. The
HERA data at low $Q^2$ shown in Fig~1.a, indicate the strong
violation of  the $\sigma_\phi/\sigma_\rho$ ratio from 2/9 value
which was found to be caused by the flavor symmetry breaking
 between $\bar u$ and $\bar s$ seas in our model. The
valence quark contribution to $\sigma_\rho$ decreases this ratio
at HERMES energies \cite{gk06}.

The model results reproduce well the energy dependence of the
$\rho$ and $\phi$ production cross section from HERMES to HERA
energies \cite{gk06}. When we extended our analysis to lower
energies $W \sim 2.5 \mbox{GeV}$, we found a good description of
the $\phi$ cross section  at CLAS \cite{clas}. This means that the
model results for gluon and sea quark, which are essential for
$\phi$ production are fine. However, for $\rho$ production, where
the valence quarks substantially contribute, we have a different
situation. At the HERMES energies the quark contribution has
maximal value  of the order of gluon and sea effects. At lower
energies the quark contribution decreases, as well as the gluon
and sea one, and the cross section falls at energies $W \le
5\mbox{GeV}$. This is in contradiction with CLAS \cite{clas}
results which show essential increase of $\sigma_\rho$ in this
energy range,
 Fig~1.b. This means that we have a problem only with the
valence quark contribution at low JLAB  energies.

\begin{figure}[h!]
\begin{center}
\begin{tabular}{cc}
\includegraphics[width=6.7cm]{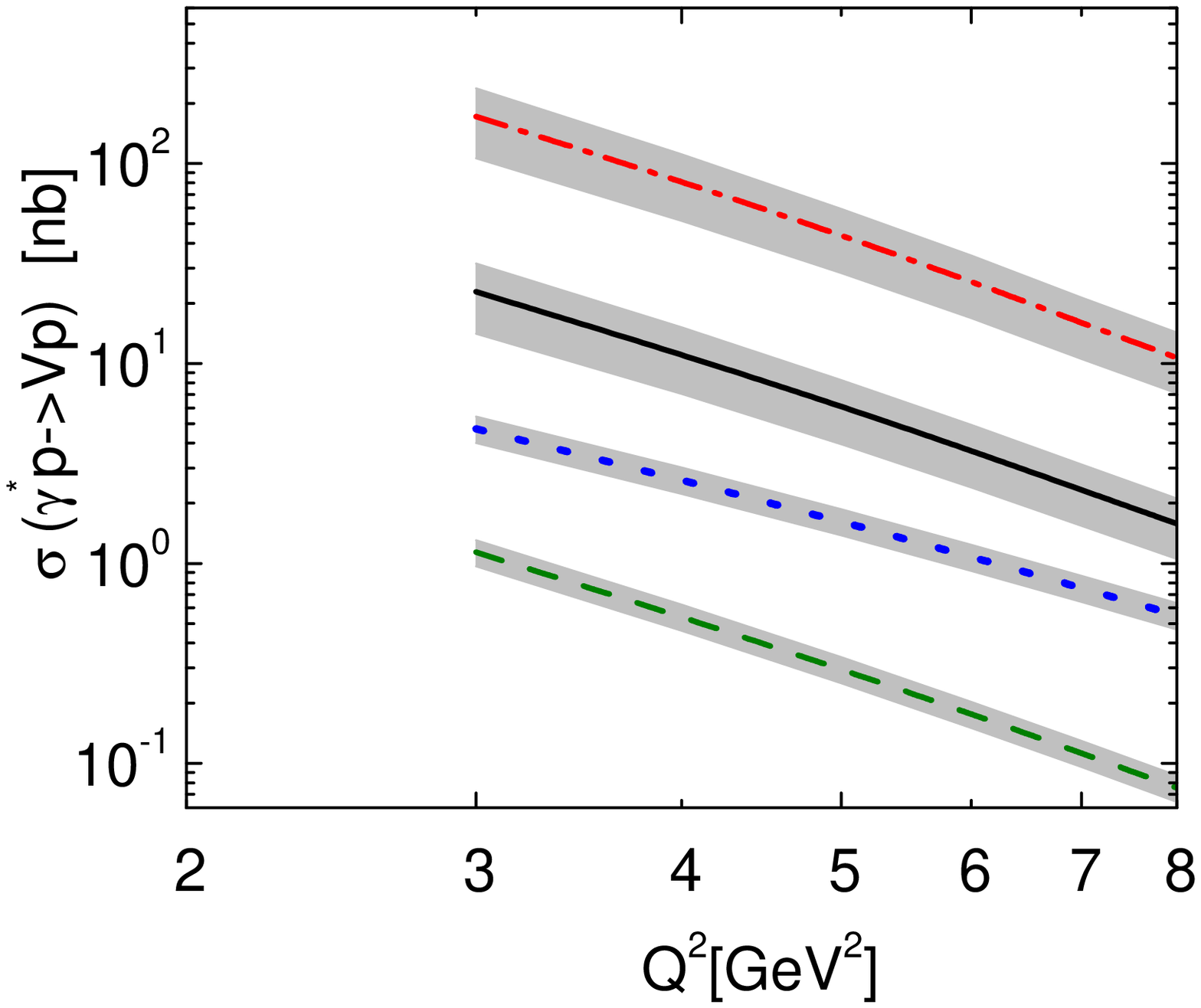}&
\includegraphics[width=6.7cm]{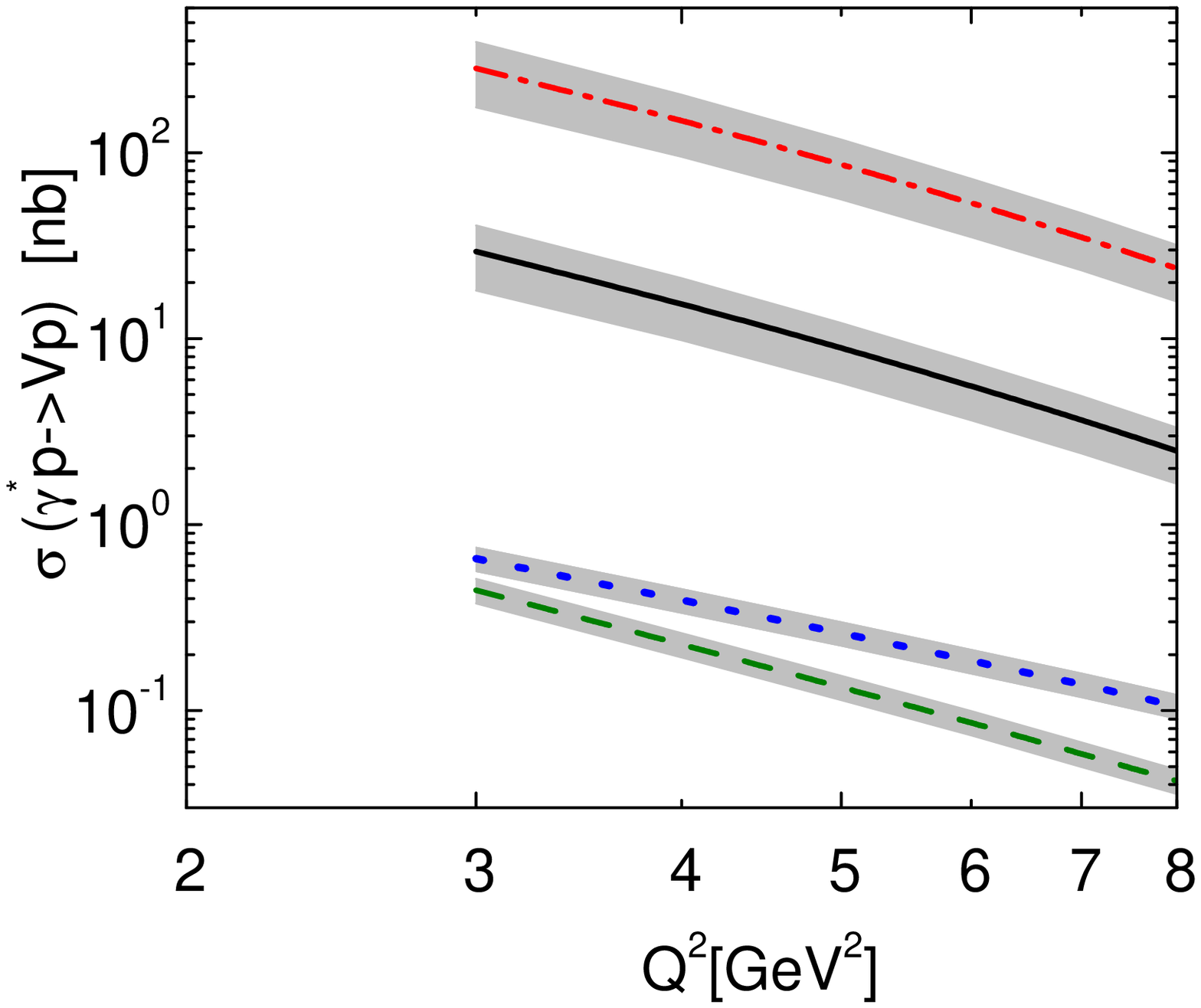}\\
{\bf(a)}& {\bf(b)}
\end{tabular}
\end{center}
\caption{ {\bf(a)} Predicted integrated over $t$ cross section
  at HERMES $W = 5\mbox{GeV}$ energies for various mesons.
    Dotted-dashed line $\rho^0$;
    full line $\omega$; dotted line $\rho^+$ and dashed line $K^{* 0}$.
     {\bf(b)} The same at COMPASS $W = 10\mbox{GeV}$ energies. }
\end{figure}

Using the combinations of quark contribution to meson production
amplitudes  (\ref{quarks}), (\ref{quarkspl}) we can predict cross
section  for various meson production. Our model results for the
 integrated over $t$ cross sections for the $\rho^0$,
$\omega$, $\rho^+$ and $K^{*0}$ electroproduction are shown in
Fig. 2.a at the energy $W = 5\mbox{GeV}$ and  in Fig. 2.b at the
energy $W = 10\mbox{GeV}$. It can be seen that the cross sections
for the $\rho^0$ and $\omega$ production increase with energy at
fixed $Q^2$ due to gluonic and sea contributions which grow with
energy for $\xi\to 0$. The cross section of $\gamma^*p\to\rho^+ p$
decreases with energy since the dominant valence quark
contributions lead to $\sigma\propto W^{4(\alpha_{\rm val}(0)-1)}$
($\alpha_{\rm val} \sim 0.5)$. This decrease for $K^{*0}$ is
smaller  than for the $\rho^+$ channel due to the strange quark
contribution which has the same energy dependence as the gluon
one.  Our predictions for the cross sections can be used to
estimate the possibility to measure asymmetries in the
corresponding processes.

The GPD $E$, which is responsible for proton helicity flip, is not
well known as yet.  We  constructed the GPD $E$ from double
distributions and constrained it by the Pauli form factors of the
nucleon \cite{pauli}, positivity bounds and sum rules. We would
like to note that the first moment of $e$ is proportional to quark
anomalous magnetic moment
\begin{equation}
E^a(x,0,0)=e^a(x);\;\;\;\;\int^1_0 dx e^a_{val}(x)=\kappa^a.
\end{equation}
The $\kappa^u$ and $\kappa^d$ have different signs. Using this
fact we can conclude that the GPD $E^u$ and $E^d$ have different
signs too. From equations (1) we see that we should have essential
compensation of $E^u$ and $E^d$ effects for $\rho$ meson and
enhancement of there contributions for $\omega$ production. The
GPD $H$ was taken from our analysis of the vector meson
electroproduction cross section.

We calculate  the $A_{UT}$ asymmetry for transversally polarized
protons in \cite{gk08}. The asymmetry is sensitive to
interferences of the amplitudes determined by the $E$ and $H$
GPDs.
\begin{equation}\label{aut}
A_{UT}= \propto \frac{\mbox{Im}<E^*>\, <H> }{|<H>|^2}.
\end{equation}
 Our results for the
$\sin(\phi-\phi_s)$ moment of the $A_{UT}$ asymmetry for the
$\rho^0$ production \cite{gk08} describe fine HERMES data
\cite{hermesaut}. Predictions  for COMPASS on $t$- dependence of
asymmetry at $W=8 \mbox{GeV}$ are shown in Fig. 3.a and are in
good agreement with preliminary COMPASS data \cite{sandacz}. As a
result of essential compensation of the GPD $E$ for valence quark,
the $A_{UT}$ asymmetry for $\rho^0$ is predicted to be quite
small.

 Predictions for the $A_{UT}$ asymmetry at $W=5
\mbox{GeV}$ and $W=10 \mbox{GeV}$  were given for the $\omega$,
 $\rho^+$, $K^{*0}$ mesons \cite{gk08}. We show our results
for $\omega$ $A_{UT}$ asymmetry at HERMES in Fig~3.b. Our
prediction for asymmetry is negative and not small. This is caused
by the fact that GPD $E^u$ and $E^d$  do not compensate each other
in this reaction. We hope that it will be possible to analyze the
$A_{UT}$ asymmetry for $\omega$ at HERMES and COMPASS despite the
smallness of the cross section for this reaction, Fig.~2.
Predictions for $\rho^+$ asymmetry is positive and rather large
$\sim$ 40\%. But smallness of the cross section does not give a
good chance to measure this asymmetry.

\begin{figure}[h!]
\begin{center}
\begin{tabular}{ccc}
\includegraphics[width=6.8cm,height=5.3cm]{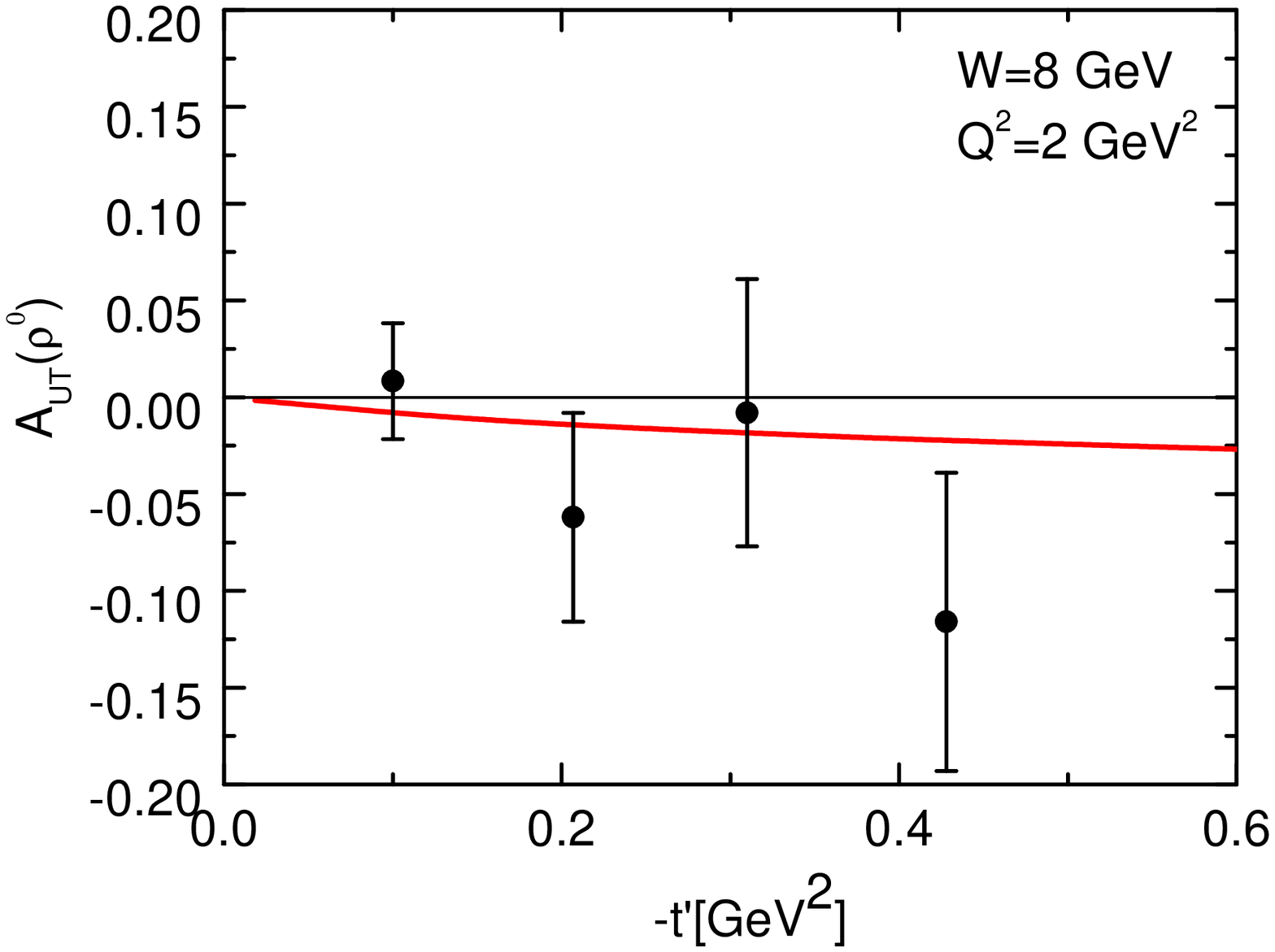}&
\includegraphics[width=6.8cm,height=5.3cm]{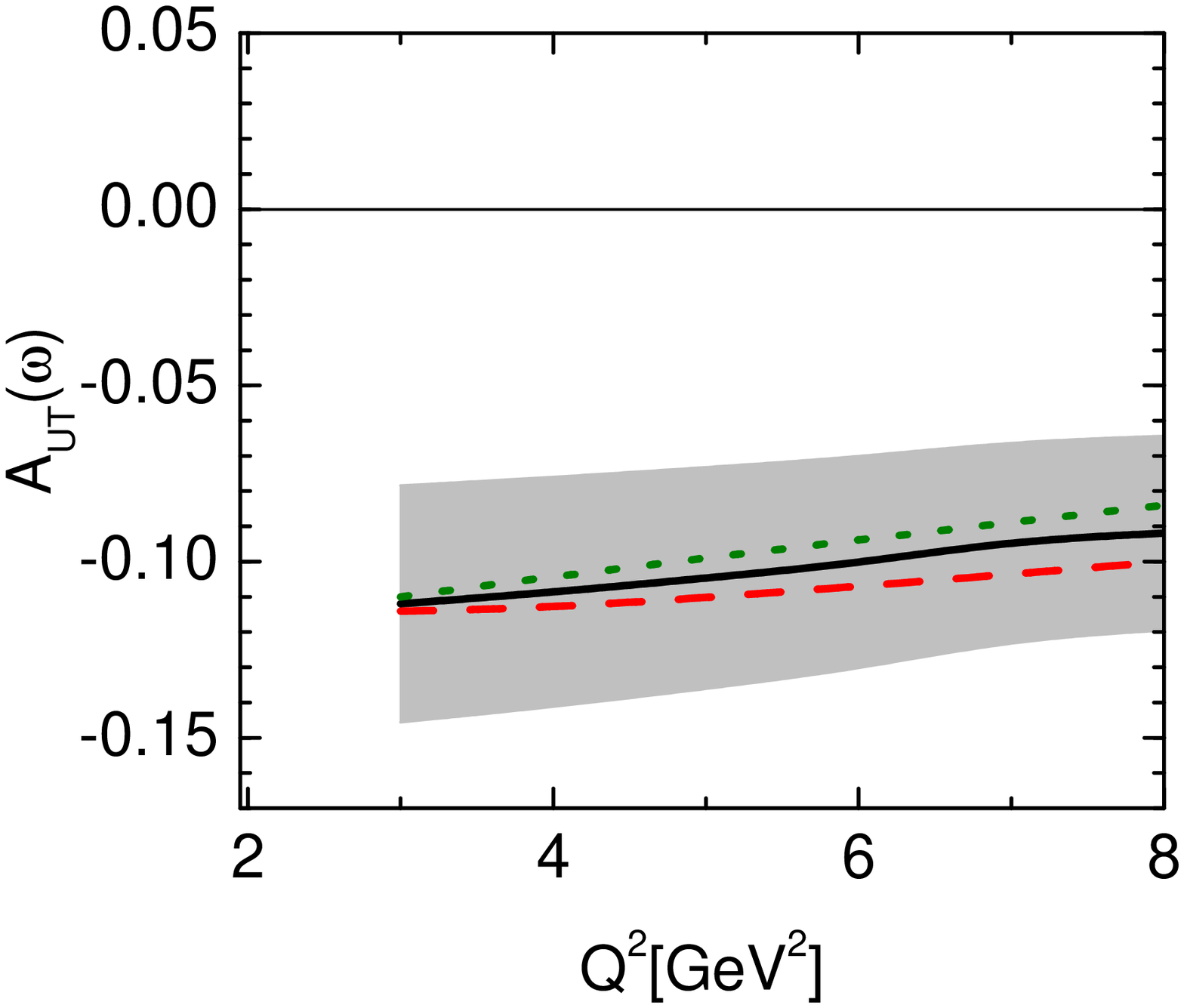}\\
{\bf(a)}& {\bf(b)}
\end{tabular}
\end{center}
\caption{ {\bf(a)}  Predictions for $A_{UT}$ asymmetry  $W=8
\mbox{GeV}$ and $Q^2=2 \mbox{GeV}^2$. Preliminary data are from
COMPASS. {\bf(b)} Predicted $A_{UT}$ asymmetry for $\omega$
production at HERMES energies}
\end{figure}
Information on the parton angular momenta can be obtained from the
Ji sum rules \cite{ji}
\begin{equation}\label{ji}
  J^q =\frac{1}{2}\int x dx (H^q(x,\xi,0)+E^q(x,\xi,0))
\end{equation}
In our model we found  not small angular momenta for $u$ quarks
and gluons
\begin{equation}
<J^u_v>=0.222, \;\;\;<J^d_v>=-0.015,\;\;\;<J^g>=0.214,
\end{equation}
which are not far from the lattice results.

\section{Electroproduction of pions}
In this process the amplitude with longitudinally polarized
photons ${\cal M}^{\pi^+}_{0\nu',0\nu}$ dominates. The amplitudes
with transversally polarized photons are suppressed as $1/Q$.
These leading twist amplitudes within the handbag approach can be
written at large $Q^2$ as  \cite{gk09}
\begin{equation}\label{pip}
{\cal M}^{\pi^+}_{0+,0+} \propto \sqrt{1-\xi^2}\,
                             \,[\langle \tilde{H}^{(3)}\rangle
  - \frac{2\xi mQ^2}{1-\xi^2}\frac{\rho_\pi}{t-m_\pi^2}];\;
{\cal M}^{\pi^+}_{0-,0+} \propto
\frac{\sqrt{-t^\prime}}{2m}\,\Big[ \xi \langle
\widetilde{E}^{(3)}\rangle +
2mQ^2\frac{\rho_\pi}{t-m_\pi^2}\Big]\,.
\end{equation}
The first terms in (\ref{pip}) represent the handbag contribution
defined in (\ref{amptt}) with
$\tilde{F}^{(3)}=\tilde{F}^{(u)}-\tilde{F}^{(d)}$ that is
determined by $p \to n$ transition GPD. The subprocess amplitudes
are calculated within the modified perturbative approach.  The
second terms in (\ref{pip}) are connected with the pion
contribution, Fig.4, which is treated with the fully
experimentally measured electromagnetic form factor of the pion.

\begin{figure}[h!]
\begin{center}
\includegraphics[width=11.0cm,height=3.cm]{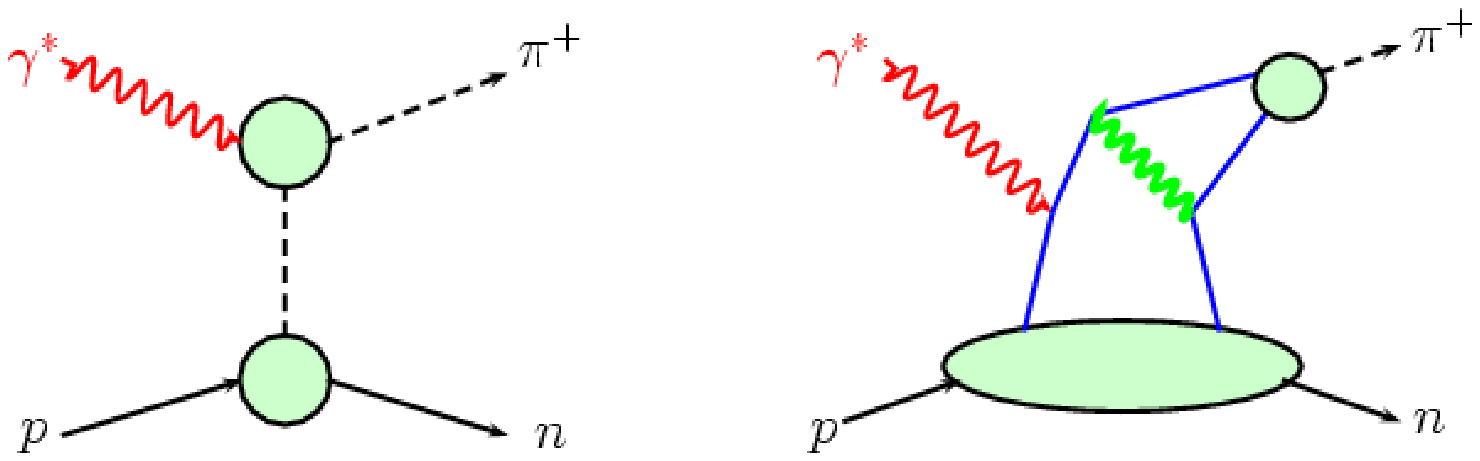}
\end{center}
\caption{ The pion pole and
 handbag contributions to $\pi^+$ production.}
\end{figure}

Using  (\ref{pip}) we calculate all amplitudes with exception of
${\cal M}_{0-,++}$. It can be found that in the handbag
approximation this amplitude should behave as $-t'$ at small
momentum transfer. However, from the  angular momentum
conservation we have the rule that $M_{\mu' \nu',\mu \nu} \propto
\sqrt{-t'}^{|\mu-\nu-\mu' +\nu'|}$ and the amplitude ${\cal
M}_{0-,++}$ should be constant at small $t'$. This problem can be
solved if we  consider  a twist-3 contribution to the amplitude
${\cal M}_{0-,++}$. In order to estimate this effect, we use a
mechanism that consists of the helicity-flip GPD $H_T$ and the
twist-3 pion wave function. The GPD $H_T$ is calculated using
double distribution and parameterization \cite{gk09} for
transversity PDFs.
\begin{figure}[h!]
\begin{center}
\begin{tabular}{cc}
\includegraphics[width=6.8cm,height=5.3cm]{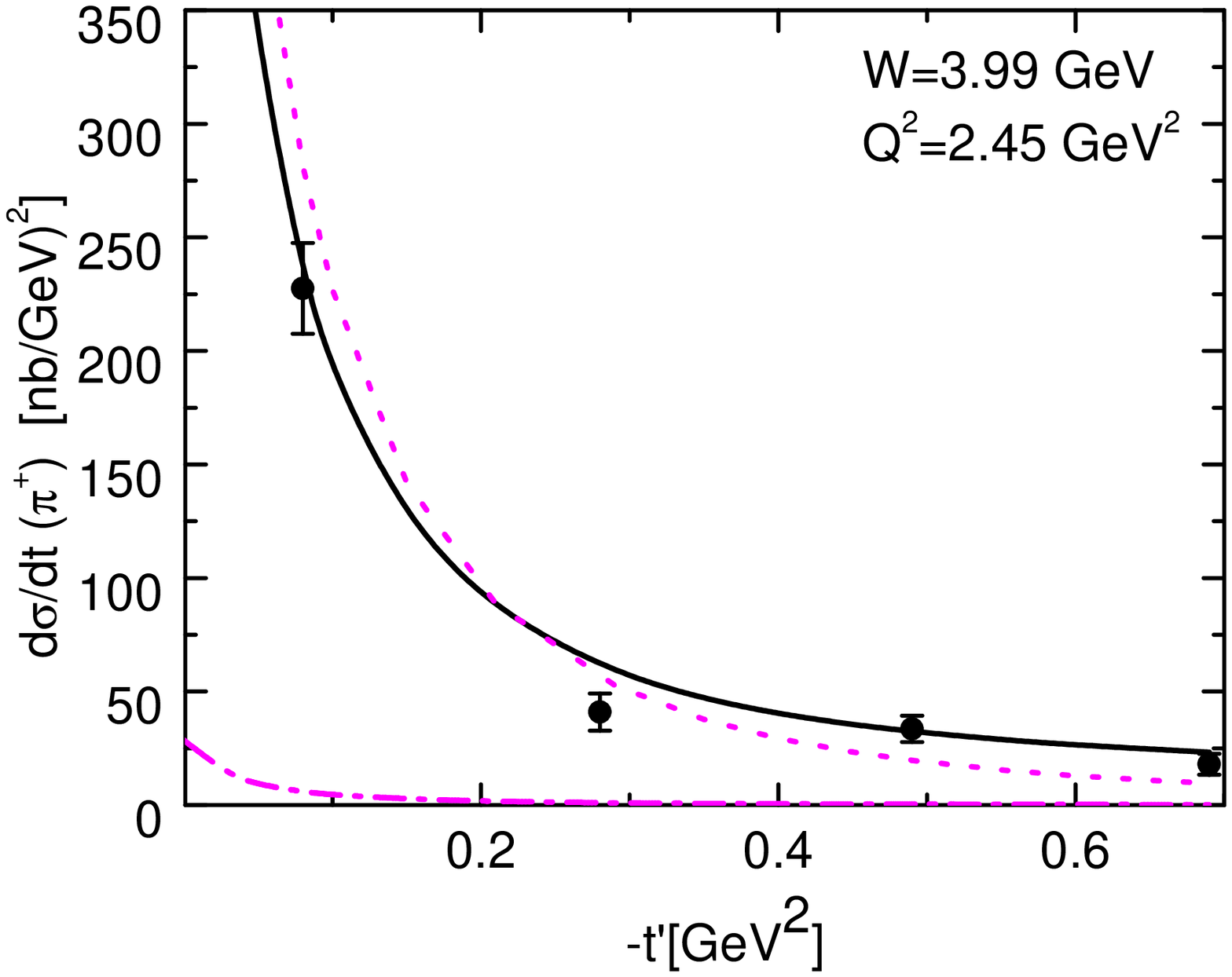}&
\includegraphics[width=6.8cm,height=5.3cm]{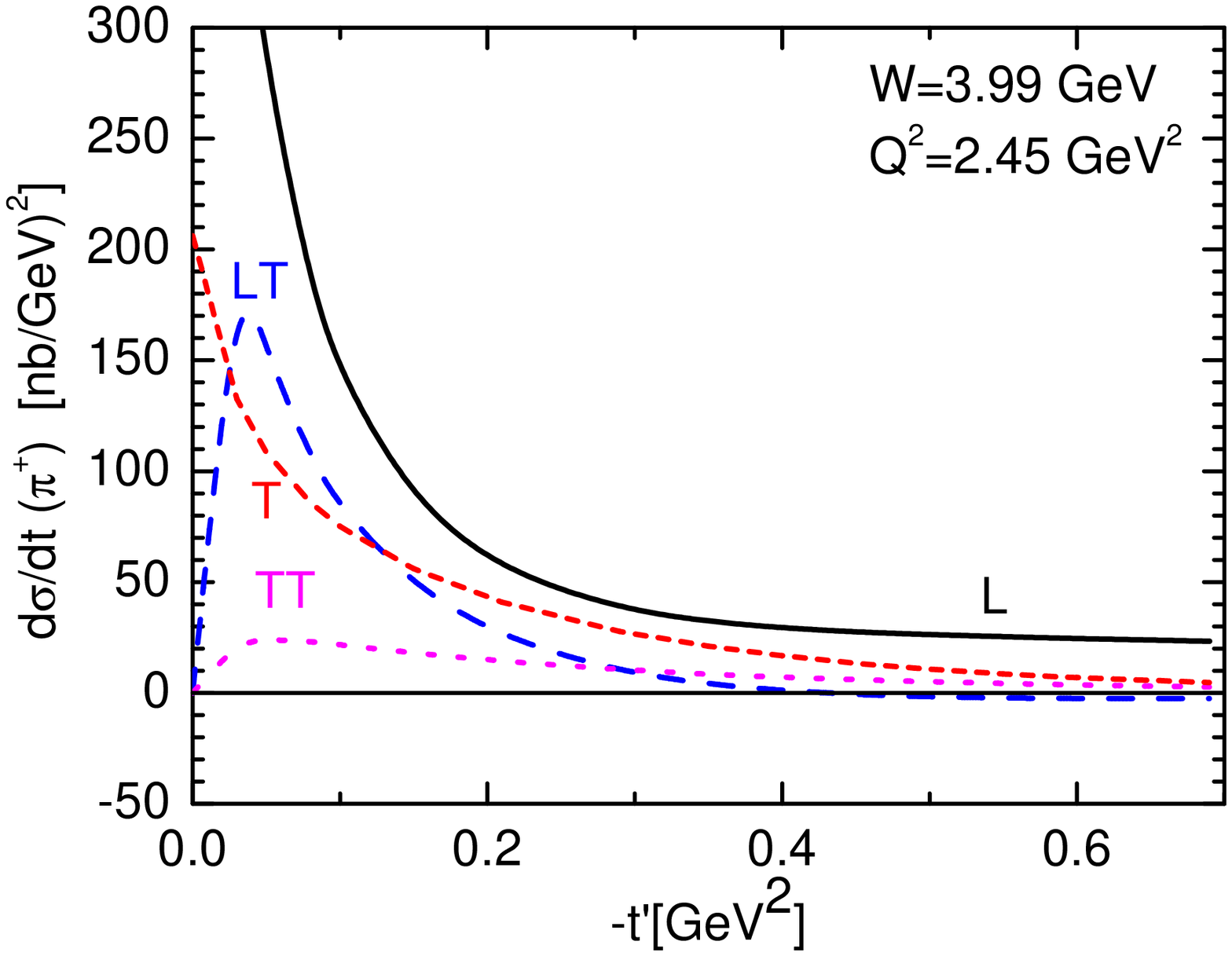}\\
{\bf(a)}& {\bf(b)}
\end{tabular}
\end{center}
\caption{ {\bf(a)} The unseparated cross section of $\pi^+$
production- solid line. Dashes and dot-dashed lines are the pion
pole contribution to unseparated and transversal cross section.
{\bf(b)} The partial cross section $d \sigma_L/dt, d \sigma_T/dt,
d \sigma_{LT}/dt, d \sigma_{TT}/dt$.}
\end{figure}

In Fig. 5.a, we show the full   cross section of $\pi^+$
production at HERMES together with the pole contribution to $d
\sigma/dt$ and  $d \sigma_T/dt$. In Fig. 5.b,  the partial cross
sections of $\pi^+$ production at HERMES are shown. The
longitudinal and full   cross sections are large at small $t'$ and
decrease rapidly with growing $-t'$. This is caused by the large
pion pole contribution to the longitudinal amplitude. The
transverse cross section is quite large because of the twist-3
effects. The pion pole contribution is rather small there.

Our results \cite{gk09} on the cross section and six moments of
spin asymmetries for the polarized target are in good agrement
with HERMES \cite{hermespi} experimental data. We show, as an
example, our results  for the $\sin{(\phi_s)}$ moment of the
$A_{UT}$ and $A_{UL}$ asymmetry in $\pi^+$ production in Fig. 6
with and without twist-3 contribution. It can be seen that twist-3
effects are very essential in description of the polarized data.

\begin{figure}[h!]
\begin{center}
\begin{tabular}{cc}
\includegraphics[width=6.8cm,height=5.3cm]{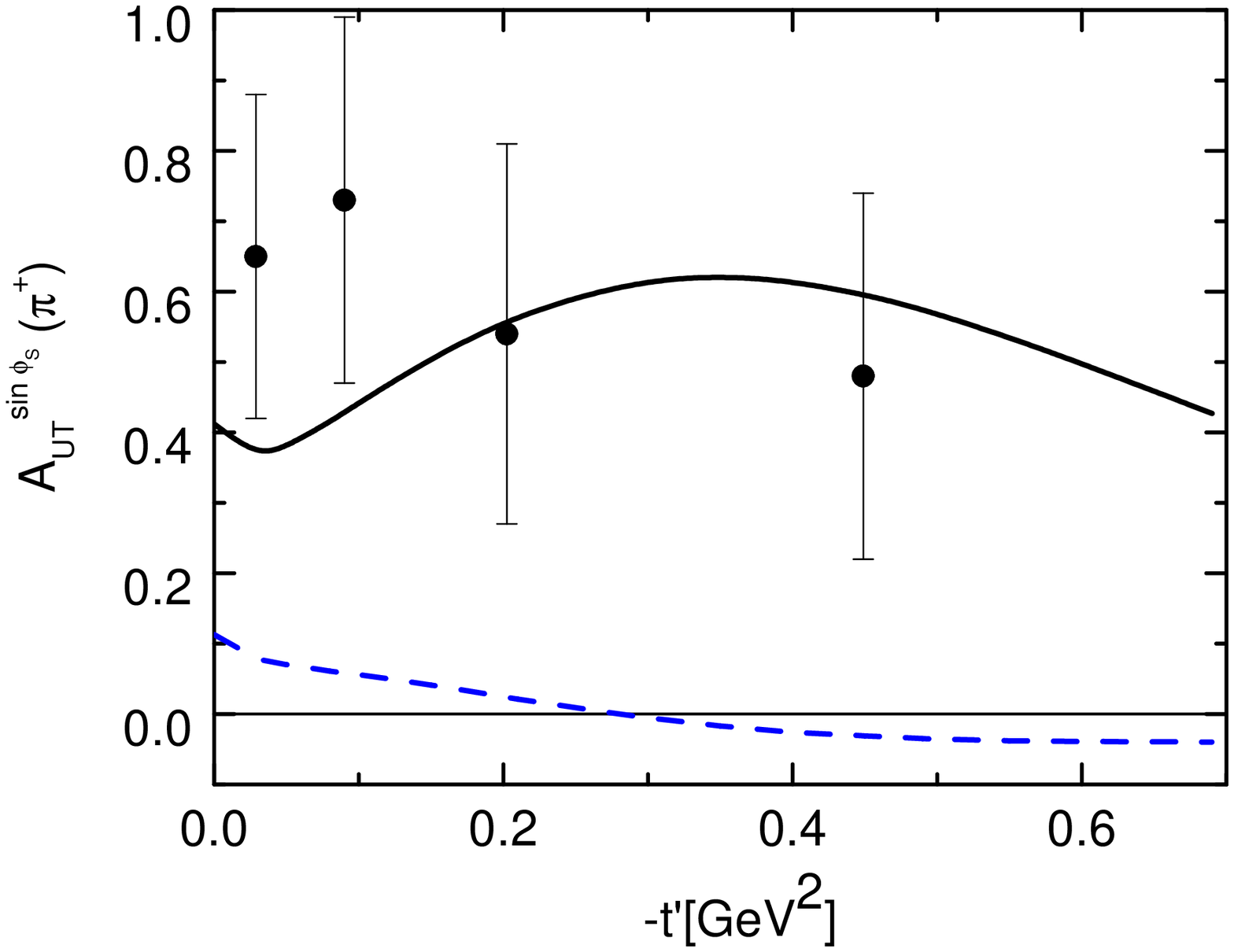}&
\includegraphics[width=6.8cm,height=5.3cm]{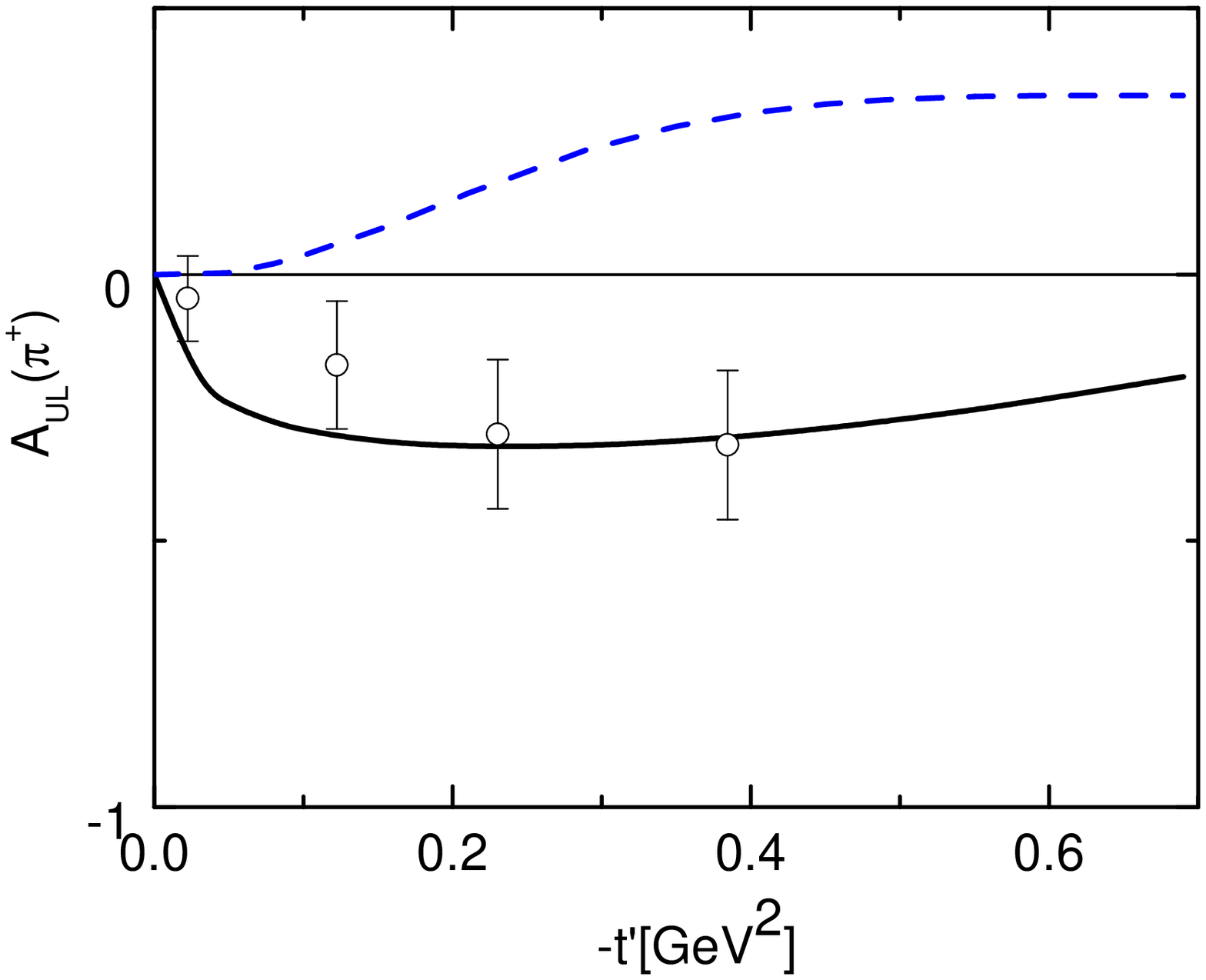}\\
{\bf(a)}& {\bf(b)}
\end{tabular}
\end{center}
\caption{ {\bf(a)} The $sin{\phi_s}$ moment of $A_{UT}$ asymmetry
at HERMES. {\bf(b)} Our results for $A_{UL}$ asymmetry at HERMES.
Dashed line obtained by neglecting the twist-3 contribution.}
\end{figure}
 \section{Conclusion and Summary}
In this report, we have studied the light vector and pseudoscalar
meson electroproduction within the handbag approach.  The
amplitude at large $Q^2$ factorizes  into a hard subprocess  and
GPDs. The model to calculate the hard subprocess amplitude is
based on MPA, where the transverse quark momenta and the Sudakov
factors were taken into account. The transverse higher twist $TT$
amplitudes with transversally polarized photons can be calculated
in the model due to the regularization of the end-point
singularities by the transverse quark momenta $\vk$. This gives a
possibility to study spin effects in the vector meson production
in our model. We describe  the cross section at quite low $Q^2$
quite well because the $k_\perp^2/Q^2$ corrections in the
propagators decrease the cross section by a factor of about 10 at
$Q^2 \sim 3\mbox{GeV}^2$ and they become close to experiment.

In the model, a good description of the cross section from HERMES
to HERA energies \cite{gk06} is observed. The gluon and sea
contributions are predominated at energies $W \geq 10 \mbox{GeV}$,
while the valence quarks are essential  at HERMES and COMPASS
energies. At lower energies  the cross section of the $\rho$ and
$\phi$ production decreases because all parton distributions go to
zero at large $x$. This is in agrement with JLAB results for the
$\phi$ production but in contradiction with essential increase of
$\sigma_\rho$ for $W<5 \mbox{GeV}$. Thus, the model describes
correctly the gluon and sea contributions up to low CLAS energies
and has  problems with valence quarks.

The model provides a good description of the spin effects,
including the $R$ ratio and SDME for the light meson production in
a wide energy range \cite{gk05,gk06,gk07q}.
 We would like to point out that study of SDME gives  important
information on different  $\gamma \to V$ transition amplitudes.
Our model predicts quite small  $\delta_{LT}$, which is in
contradiction with the experimental results which show  a quite
large phase difference $\delta_{LT} \sim 20-30^o$ \cite{hermes}.
Further theoretical and experimental investigations are necessary
to shed  light on the problems with $\rho$ cross section growing
at low $W$ and the phase shift between the $LL$ and $TT$
amplitudes.

Using the GPD $E$ estimation on the basis of the nucleon Pauli
form factor we predict the cross sections and $A_{UT}$ asymmetries
for the $\rho^0$, $\omega$, $\rho^+$ and $K^{*0}$
electroproduction \cite{gk08}. These reactions  give a possibility
to test the $H$ and $E$ GPDs for valence and sea quarks. The
experimental data which are available only for the $\rho^0$
production are described well. We predict large negative $A_{UT}$
asymmetry for $\omega$ production which most probably may be
studied at HERMES and COMPASS.

Our analysis on pion production \cite{gk09} gives an access to
$\tilde H$ and $\tilde E$ GPDs. The essential twist -3 effects in
the ${\cal M}_{0-,++}$ amplitude were found, which is very
important in understanding  the cross section and spin asymmetries
in $\pi^+$ production. Further theoretical and experimental
studies of these precesses are needed, especially for $\pi^0$
production which was analyzed for the first time in \cite{gk09}.
Thus, we can conclude that the gluon and various quark GPDs can be
probed in the meson electroproduction.

 \bigskip

 This work is supported  in part by the Russian Foundation for
Basic Research, Grant  09-02-01149  and by the Heisenberg-Landau
program.

\end{document}